\documentstyle[twocolumn,seceq,epsf]{jpsj}

\def\runtitle{Magnetization plateaus in polymerized $S=1/2$ chains
}
\def\runauthor{W. {\sc Chen}, K. {\sc Hida} and B. C. {\sc Sanctuary}}

\title
{
Magnetization plateaus in antiferromagnetic-(ferromagnetic)$_{n}$ polymerized $S=1/2$ XXZ chains
}

\author
{
Wei {\sc Chen}\footnote{E-mail: chen@chemistry.mcgill.ca, Present address: Department of Chemistry, McGill University, Montreal, PQ, Canada H3A 2K6 
}, Kazuo {\sc Hida}\footnote{E-mail: hida@phy.saitama-u.ac.jp} and B. C. {\sc Sanctuary}$^1$
}

\inst
{
Department of Physics, Saitama University, Urawa, Saitama, Japan 338-8570\\
$^1$ Department of Chemistry, McGill University, Montreal, PQ, Canada H3A 2K6 
}

\recdate
{
\today
}

\abst
{The plateau-non-plateau transition in the antiferromagnetic-(ferromagnetic)$_{n}$ polymerized $S=1/2$ XXZ chains under the magnetic field is investigated. The universality class of this transition belongs to the Brezinskii-Kosterlitz-Thouless (BKT) type. The critical points are determined by level spectroscopy analysis of the numerical diagonalization data for $4 \leq  p \leq 13$ where $p(\equiv n+1)$ is the size of a unit cell. It is found that the critical strength of ferromagnetic coupling decreases with $p$ for small $p$ but increases for larger enough $p$. It is also found that the plateau for large $p$ is wide enough for moderate values of exchange coupling so that it should  be easily observed experimentally. This is in contrast to the plateaus for $p = 3$ chains which are narrow for a wide range of exchange coupling even away from the critical point.
}

\kword
{
magnetization plateau, level spectroscopy, exact diagonalization, BKT transition, twisted boundary condition
}

\begin{document}
\sloppy
\maketitle

\section{Introduction}
Magnetization plateaus in one dimensional Heisenberg chains are attracting much attention as the field induced spin gap states. On the plateau, some spins are partly quenched by the magnetic field while the remaining spins form a spin gap state.\cite{hk,ts2,okamoto,tone,mo,to2,to,na,ak,kiok,kiok1,naru,WS} It is evident that such magnetization plateaus can be realized in various non-homogeneous Heisenberg chains. Oshikawa, Yamanaka and Affleck\cite{mo} proposed the  necessary condition for the magnetization plateaus as $p(S-m^{z})=q\equiv \mbox{integer}$ where $p$ is the periodicity of the magnetic ground state in the thermodynamic limit, $S$ is the magnitude of the spin, and $m^{z}$ is the magnetization per site. 

In this context, the spin chains consisting of periodic arrays of ferromagnetic and antiferromagnetic bonds have been widely investigated. The $S=1/2$ ferromagnetic-antiferromagnetic alternating chain has a spin gap ground state\cite{hk0,hki,hkii} for arbitrary values of coupling constants. This state can be regarded as the plateau state with zero magnetization ($p=2, S=1/2, q=1$ and $m^z=0$). In the isotropic case, this energy gap survives as the well-known Haldane gap even if the strength of the  ferromagnetic bonds $j$ is infinite.\cite{hk0} 

There appears a  magnetization plateau in the magnetization curve of the $S=1/2$ ferromagnetic-ferromagnetic-antiferromagnetic trimerized chain at 1/3 of the saturation magnetization ($p=3, S=1/2, q=1$ and $m^z=1/6$).\cite{hk} In this case, Kitazawa and Okamato\cite{ak} have shown that the plateau vanishes by the Brezinskii-Kosterlitz-Thouless (BKT) type transition for $j \geq j_c \simeq 15.4$ where the energy unit is the strength of the antiferromagnetic bonds.

In the present work, we investigate the plateau-non-plateau transition for general values of $p$ using the method of ref. \citen{ak}. Intuitively, the quantum fluctuations are expected to be suppressed as the number of the ferromagnetic bonds increases. Therefore it is natural to expect that the magnetization plateau, which is an essentially quantum phenomenon, becomes less favourable as the number of the ferromagnetic bonds increases. Surprisingly, our numerical calculation shows that this is {\it incorrect}. We discuss the  physical origin of this unexpected behavior.

This paper is organized as follows. In the next section, the model Hamiltonian is defined and the numerical results of the magnetization curve and the phase diagram of the plateau-non-plateau transition  is presented. The physical reason of the enhancement of the plateau state for large $p$ is also explained in this section. The final section is devoted to a summary and discussion.

\section{Numerical Results}
\subsection{Model Hamiltonian}
The Hamiltonian of the antiferromagnetic-(ferromagnetic)$_{n}$ polymerized $S=1/2$ XXZ chain in the magnetic field is given by
\begin{eqnarray}
\label{model}
{\cal H}&=&-j\sum_{l=1}^{L} \sum_{i=0}^{n-1} {\cal H}_{(n+1)l+i}(1)+\sum_{l=1}^{L}{\cal H}_{(n+1)l+n}(\Delta) \nonumber \\
&-&g\mu_{\rm B}H\sum_{l=1}^{(n+1)L}S_{l}^{z}.
\end{eqnarray}
where 
\begin{equation}
{\cal H}_l(\Delta)=S_{l}^{x}S_{l+1}^{x}+S_{l}^{y}S_{l+1}^{y}+\Delta S_{l}^{z}S_{l+1}^{z},
\end{equation}
The ferromagnetic coupling constant, number of the ferromagnetic bond and anisotropy parameter of the antiferromagnetic bond are represented by $j$, $n$ and $\Delta$, respectively. The ferromagnetic bonds are assumed to be isotropic. The magnetic field, the electronic $g$-factor and  Bohr magneton are denoted by $H, g$ and $\mu_{\rm B}$, respectively. This model has spatial periodicity $p \equiv n+1$.  In the following, we take $g\mu_{\rm B}=1$ as unit.

In the limiting case of $j \rightarrow \infty$, this model tends to the $S=p/2$ uniform antiferromagnetic XXZ chain, so that no plateau is expected. Therefore the BKT type plateau-non-plateau transition is expected to take place at a finite critical value of $j$. 
\subsection{Magnetization plateaus}
\begin{figure}
%%\figureheight{7cm}
\epsfxsize=60mm 
\centerline{\epsfbox{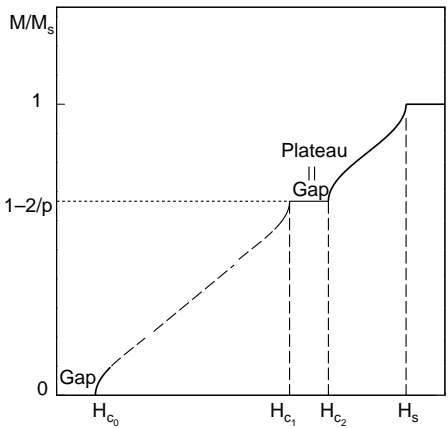}}
\caption{Schematic magnetization curve of the present model.}
\label{shem}
\end{figure}

 The schematic magnetization curve of the present model is shown in Fig. \ref{shem}. In the present paper, we concentrate on the highest plateau with $q=1$ and $m^{z}=1/2-1/p$. The total magnetization $M \equiv m^z$ on the plateau is given by $M = M_p \equiv pL(1/2-1/p)=M_s(1-2/p)$ where $M_s \equiv pL/2$ is the saturation magnetization.
\vspace{3mm}
In Fig. \ref{shem} the critical fields $H_{c0}$ corresponds to the energy gap in the absence of the magnetic field, $H_{c1}$, to the energy gap between the lowest energy with $M^{z}=M_p$ and that with $M^{z}=M_p-1$, $H_{c2}$, the energy gap between the lowest energy with $M^{z}=M_p+1$ and that with $M^{z}=M_p$. Finally $H_{s}$ is the saturation field. The lower part of the magnetization curve is represented by the dashed line where lower plateaus may appear.

\begin{figure}
%\figureheight{7cm}
\epsfxsize=70mm 
\centerline{\epsfbox{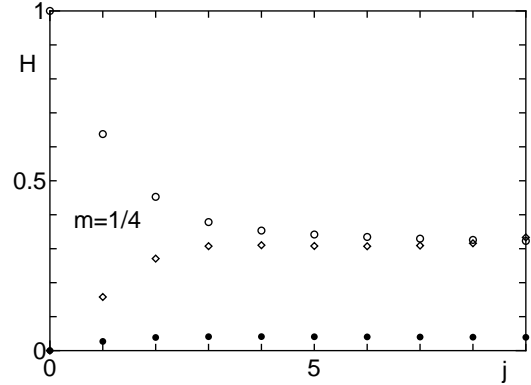}}
\caption{The critical fields for $p=4$ and $\Delta=1$. The critical fields $H_{c0}, H_{c1}$ and $H_{c2}$ are represented by $\bullet$, $\Diamond$ and $\circ$.}
\label{pla}
\end{figure}

We use an exact diagonalization method to calculate the magnetization curve. As an example, the $j$-dependence of the 3 critical fields $H_{c0}$, $H_{c1}$ and $H_{c2}$ for $p=4$ and $\Delta=1$ are shown in Fig. \ref{pla} by the symbols $\bullet$, $\Diamond$ and $\circ$, respectively, which are obtained by the Shanks'\cite{shanks} extrapolation of the results of the Lanczos exact diagonalization for $4L=8,16$ and 24 with periodic boundary condition. The plateau seems to close at a finite value of $j$. Because the non-plateau state corresponds to the gapless spin liquid and the plateau state to the spin gap state, this transition is expected to be the BKT-type transition. In the next subsection. we determine the critical point of the plateau-non-plateau transition from numerical diagonalization data for general values of $p$. 
\subsection{BKT critical point}

It is difficult to estimate the  BKT critical point precisely from standard finite size analysis of the numerical calculation data. To circumvent this difficulty Nomura and Kitazawa\cite{no3} proposed to use level spectroscopy method\cite{nomura} with twisted boundary condition. This method has been successfully applied to the ferromagnetic-ferromagnetic-antiferromagnetic trimerized Heisenberg chain by Kitazawa and Okamoto.\cite{ak} In the present work, we employ this method for the present model.  Here we do not explain how and why this method works, because methodological details are well described in ref. \citen{ak}.

The finite size critical point is determined from the crossing point of $\Delta E_{0,2}$ with the lower of $\Delta E^{c}_{1/2,0}$ or $\Delta E^{s}_{1/2,0}$ defined by,
\begin{eqnarray}
\Delta E_{0,2} &=&\frac{1}{2}\Big\{E_{0}(M_{\rm p}+2,0,1)+E_{0}(M_{\rm p}-2,0,1)\Big\} \nonumber \\
&-&E_{0}(M_{\rm p},0,1).
\end{eqnarray}
and  
\begin{equation}
\Delta E_{1/2,0}^{c}=E^{TBC}(M_{\rm p},1)-E(M_{\rm p},0,1)
\end{equation}
\begin{equation}
\Delta E_{1/2,0}^{s}=E^{TBC}(M_{\rm p},-1)-E(M_{\rm p},0,1)
\end{equation}
where $E_{0}(M^{z},k,P)$ is the lowest energy under a periodic boundary condition with magnetization $M^{z}$, wave number $k$ and parity $P$. The energy  $E^{TBC}(M^{z},P)$ is the lowest energy with the twist boundary conditions with  magnetization $M^{z}$ and parity $P$.

\begin{figure}
%\figureheight{7cm}
\epsfxsize=70mm 
\centerline{\epsfbox{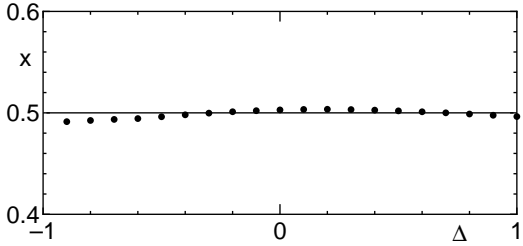}}
\caption{The extrapolated values of the averaged scaling dimension $x$ on the critical points for $p=4$. The solid line is $x=0.5$}
\label{x}
\end{figure}

To confirm the reliability of this method the following average 
\begin{equation}
\label{ve}
x=\frac{x_{1/2,0}^{c}(L)+3x_{1/2}^{s}(L)}{4}.
\end{equation}
is close to 0.5 at the critical points\cite{ak} where $x_{1/2,0}^{c,s}(L)$ are the scaling exponents corresponding to  $E_{1/2,0}^{c,s}$ and defined by
\begin{equation}
x_{1/2,0}^{c,s}=\frac{L}{2\pi v_{\rm s}}\Delta E_{1/2,0}^{c,s}.
\end{equation}
Here $v_{s}$ is the spin wave velocity given by
\begin{equation}
v_{\rm s}=\lim_{L\rightarrow \infty} \frac{L}{2\pi}[E_{M_p, k_{1}}(L)-E_{M_p}(L)],
\end{equation}
where $E_{M_p, k_{1}}(L)$ is the energy of the excited state with wave number $k_{1}=\frac{2\pi}{L}$ and $M^{z}=M_{\rm p}$. The results are shown in Fig. \ref{x} for $p=4$ which confirms that $x=0.5$ holds with good accuracy.

\begin{figure}
%\figureheight{7cm}
\epsfxsize=70mm 
\centerline{\epsfbox{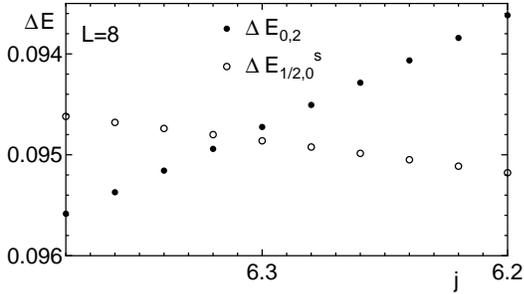}}
\caption{The $j$-dependence of the energies $\Delta E_{0,2}$ and $\Delta E_{1/2,0}^{s}$ represented by $\circ$ and $\bullet$, respectively, for $p=4, \Delta=1$ and $L=6$.}
\label{twen}
\end{figure}

Figure \ref{twen} shows the behavior of $\Delta E_{0,2}$ and $\Delta E_{1/2,0}^{s}$ for $L=6$ unit cells and $p=4$. From the crossing point, we obtain $j_{c}(L)=6.310$ for $L=6$ unit cells. The BKT transition point for the infinte system can be obtained by extrapolating from $L=2, 4$ and 6 to $N\rightarrow \infty$ as $j_{c}=6.260$ assuming the extrapolation formula $j_{c}(L)=j_{c}+\frac{c_{1}}{L^{2}}+\frac{c_{2}}{L^{4}}$.


\begin{figure}
%\figureheight{7cm}
\epsfxsize=70mm 
\centerline{\epsfbox{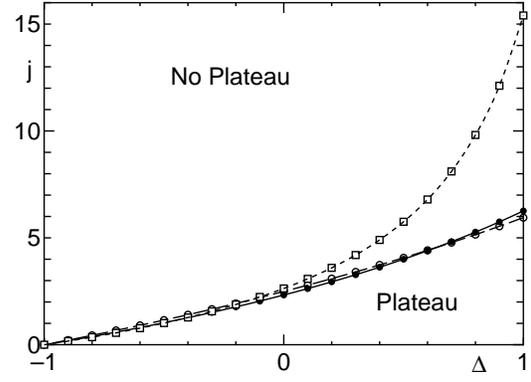}}
\caption{The phase boundaries on the $\Delta-j$-plane for $p=3, 4$ and 5 represented by $\Box$ $\bullet$ and $\circ$ , respectively. The lines are fitted.}
\label{phase}
\end{figure} 

\begin{figure}
%\figureheight{7cm}
\epsfxsize=70mm 
\centerline{\epsfbox{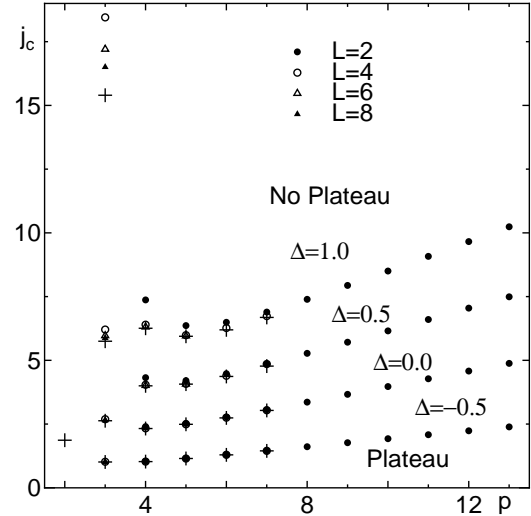}}
\caption{The $p$-dependence of the critical point $j_{c}$ for $\Delta=1.0, 0.5, 0.0$ and $-0.5$ from top to bottom. The  different symbols distinguish $j_c$'s for different system sizes $L$. The $+$ symbols are the values of $j_c$ extrapolated to $L \rightarrow \infty$. The result for $p=3$ is taken from ref. \citen{ak}. For $p=2$, $j_c \rightarrow \infty$ for $\Delta \geq 0$.\cite{hki,hkii}}
\label{jdep}
\end{figure}

The phase boundaries are shown in Fig. \ref{phase} for $p=3, 4$ and 5 by $\Box$, $\bullet$ and $\circ$, respectively. Extrapolation is made from $L=2, 4$ and 6 for $p=4$ and 5. The results for $p=3$ are taken from ref. \citen{ak}. 
From this figure, we see that the critical points $j_{c}$ does  not decrease monotonically with an increase of periodicity $p$. Therefore we further calculate the cases of longer periodicity up to $p=13$. The $p$ dependence of $j_c$ is shown in Fig. \ref{jdep} for $\Delta=1.0, 0.5, 0.0$ and $-0.5$. The different symbols represent the critical points for different system size $L$. The data for $p=2$ is also included for $\Delta=-0.5$. When $\Delta \geq 0$ and $p=2$ then $j_c$ equals infinity.\cite{hki,hkii} For $p= 4$ and 5, the calculations are carried out for $L=2, 4$ and 6 and for $p=6$ and 7,  $L=2$ and 4. The extrapolation to $L \rightarrow \infty$ is made for these cases. Nevertheless, the $L$-dependence of the critical point becomes weaker as the periodicity $p$ becomes larger as shown in Fig. \ref{jdep}. Therefore for $p \geq 8$, we can safely estimate approximate values of the thermodynamic critical point $j_c$ from that for $L=2$.  Surprisingly, the $p$-dependence of the critical point is strongly non-monotonic. The decrease of $j_c$ with $p$ for small values of $p$ is consistent with intuition that the increasing the number of the ferromagnetic bonds suppresses the quantum fluctuation. The magnetization plateau, which is the quantum origin, becomes less favoured. 

\begin{figure}
%\figureheight{4cm}
\epsfxsize=60mm 
\centerline{\epsfbox{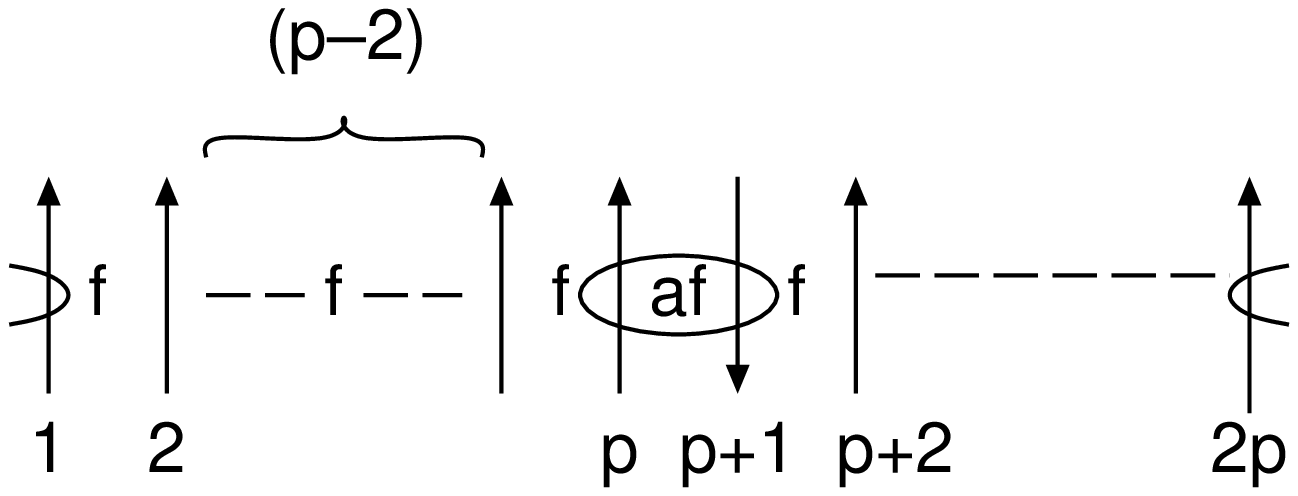}}
\begin{center}
(a)
\end{center}
\epsfxsize=60mm 
\centerline{\epsfbox{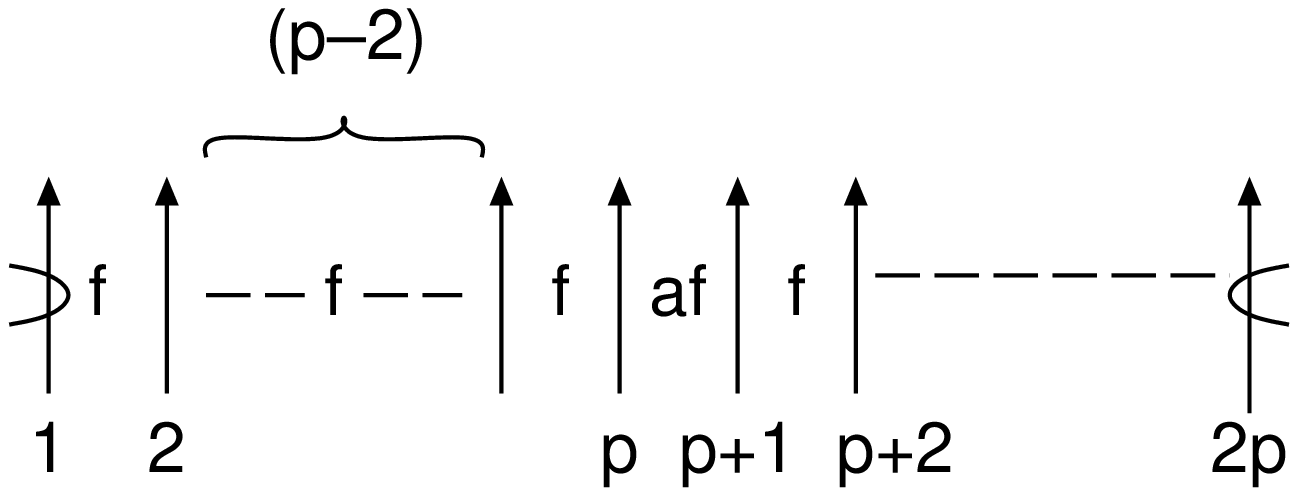}}
\begin{center}
(b)
\end{center}
\epsfxsize=60mm 
\centerline{\epsfbox{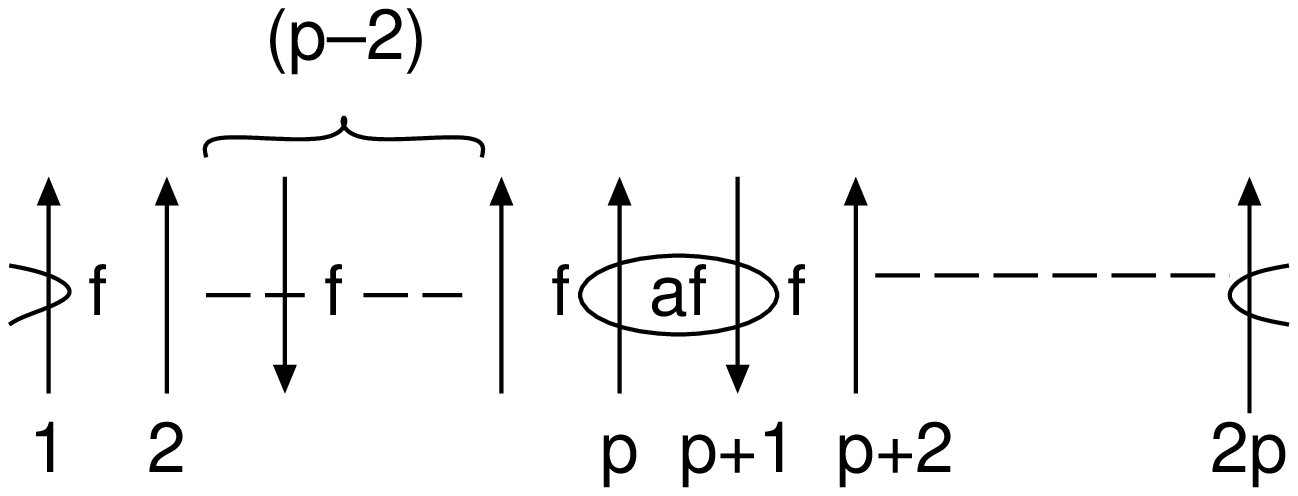}}
\begin{center}
(c)
\end{center}
\caption{The spin configurations (a) on the magnetization plateau $M^z=M_{p}$, (b) at the magnetization $M^z=M_{p}+1$ and (c) at the magnetization $M^z=M_{p}-1$. The spin pairs connected by ovals form local singlets.}
\label{mpla}
\end{figure}

\begin{figure}
%\figureheight{7cm}
\epsfxsize=70mm 
\centerline{\epsfbox{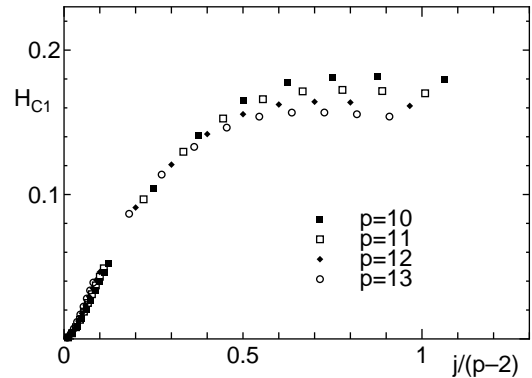}}
\caption{The $\frac{j}{p-2}$ dependence of $H_{c1}$ for $p=10, 11, 12$ and 13 from top to bottom, $H_{c1}$ scales as $\frac{j}{p-2}$.}
\label{pljop}
\end{figure}

\begin{figure}
%\figureheight{7cm}
\epsfxsize=70mm 
\centerline{\epsfbox{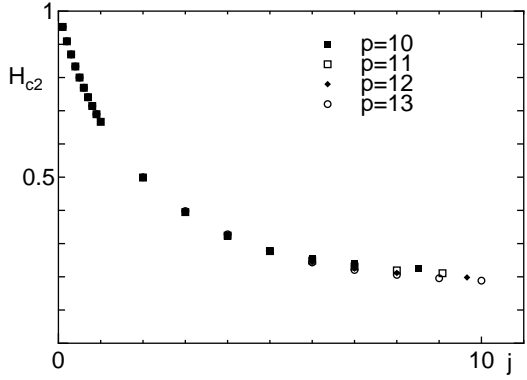}}
\caption{The $j$-dependence of $H_{c2}$ for $p=10, 11, 12$ and 13. $H_{c2}$ is almost independent of $p$ and the tends to a constant value for large $j$.}
\label{phjp}
\end{figure}

\begin{figure}
%\figureheight{7cm}
\epsfxsize=70mm 
\centerline{\epsfbox{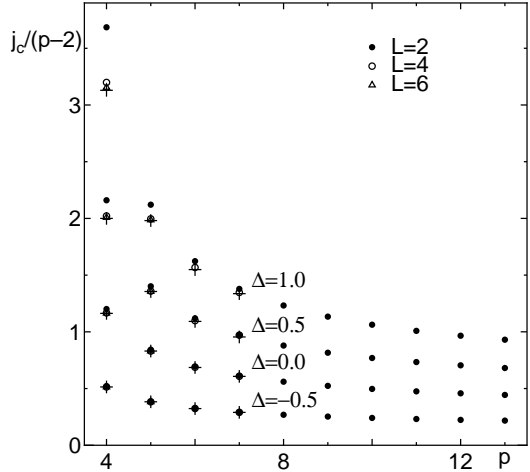}}
\caption{The $p$-dependence of $j_{c}/(p-2)$.}
\label{jop2}
\end{figure}

With the further increase of $p$, however, $j_c$ increases again almost linearly with $p$. This phenomenon can be explained by the following physical picture. On the magnetization plateau $M^z=M_{p}$, the singlet pairs are located on the antiferromagnetic bonds and other spins are alligned in the direction of the magnetic field. This is schematically shown in Fig. \ref{mpla}(a). If the magnetization becomes $M^{z}=M_{p}+1$, one singlet pair is broken as shown in Fig. \ref{mpla}(b). The energy difference $E_{0}(M_p+1)-E_{0}(M_p)$ determines the upper edge of the plateau $H_{c2}$.  Because the additional up spin is localized around the antiferromagnetic bonds, the critical field $H_{c2}$ tends to a constant value of the order of the antiferromagnetic coupling constant $\sim {\cal O}(1)$ as $p$ increases.  On the other hand, if the magnetization becomes  $M^{z}=M_{p}-1$, one of $p-2$ spins in the ferromagnetic segments is inverted as shown in Fig. \ref{mpla}(c). In the absence of coupling to the 2 boundary spins forming a local singlet, the segment of the remaining $p-2$ spins, which contain one inverted spin, form a state with total spin $S=\frac{p-2}{2}$ and magnetization $S^{z}=\frac{p-2}{2}-1$. Since the ferromagnetic bonds are isotropic this state has the same energy as the fully polarized state with $S=\frac{p-2}{2}$ and $S^{z}=\frac{p-2}{2}$. The energy change  $E_{0}(M_p)-E_{0}(M_p-1)$ is due to the interaction of the inverted spin with the singlet pair on the antiferroamgnetic bonds at the both ends of the ferromagnetic segment. The probability that the inverted spin stays on the boundary sites of the segment is proportional to $\frac{1}{p-2}$. Hence, the  energy diference $E_{0}(M_p)-E_{0}(M_p-1)$, which determines the lower edge of the plateau $H_{c1}$ is estimated as $O(\frac{j}{p-2})$ for large enough $p$. Because the plateau-non-plateau transition takes place around the point $H_{c1} \sim H_{c2}$, we find $j_{c} \propto p-2$ for large enough $p$. This explains why $j_{c}$ increases linearly with $p$ for large $p$. To substantiate this argument, $H_{c1}$ and $H_{c2}$ are calculated  numerically by exact diagonalization for $L=2$ and $\Delta=1.0$.  Figure \ref{pljop} shows that $H_{c1}$ scales with $\frac{j}{p-2}$ for $p=10,11,12$ and 13 from top to bottom although there is some deviation as the critical point is approached. Figure  \ref{phjp} shows $H_{c2}$ is approximately independent of $p$ and tends to a constant value for large $j$. These results show that the increase of the critical point for large $p$ is a consequence of delocalization of the down spin in a wide region of length $p$.  In  Fig. \ref{jop2}, $\frac{j_{c}}{p-2}$ is plotted against $p$. It is evident that $\frac{j_{c}}{p-2}$ tends to a constant value for large enough $p$.

\section{Summary and Discussion}

The magnetization plateaus at $m^{z}=1/2-1/p$ in the $S=1/2$ antiferromagnetic-(ferromagnetic)$_n$ polymerized XXZ chains are investigated by exact diagonalization of finite size systems. The plateau-non-plateau BKT transition points are determined precisely by the level spectroscopy method with twisted boundary condition.\cite{no3,ak} 

The $p$-dependence of the critical point is not a monotonic function. For small $p$, the critical point decreases with $p$ in agreement with the simple-minded intuition that an increase of the number of the ferromagnetic bonds suppresses the quantum fluctuation and the magnetization plateau, which is the quantum origin, becomes less favoured. With a further increase of $p$, however, $j_c$ increases again. This can be explained by  the delocalization of down spin within the ferromagnetic segment with length $p-2$. 

\begin{figure}
%\figureheight{7cm}
\epsfxsize=70mm 
\centerline{\epsfbox{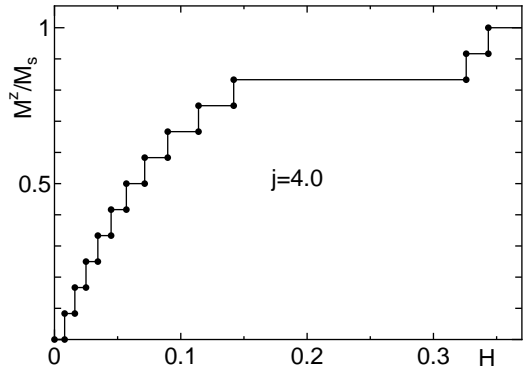}}
\caption{The magnetization curve for $p=12, \Delta=1$ and $j=4$ obtained by exact diagonalization with $L=2$.}
\label{epla12}
\end{figure}

The plateau for $p=3$ is narrow for a wide range of $j \leq j_c$.\cite{hk} This is the reason why so far this simplest example of the magnetization plateau has not even observed experimentally. Actually, no visible plateau is observed in the magnetization curve of 3CuCl$_2 \cdot$ 2dx which has $j \sim 4.5 << j_c$.\cite{hk,ajiro} In contrast, the magnetization plateaus for large $p$ is fairly wide compared to that for $p = 3$ for moderate values of $j$. As an example, we show the magnetization curve for  $p=12, \Delta=1$ and $j=4.0$ obtained by exact diagonalization for $L=2$ in Fig. \ref{epla12}. It is evident that this plateau is wide enough for experimental measurement, although the value of $j_c$ for $p=12$ is less than that for $p=3$. Thus a systematic synthesis of materials with broad spatial periodicity is hoped to experimentally confirm our predictions.

Chen and Hida thank K. Okamoto for fruitful discussion. We also thank A. Kitazawa and K. Okamoto for supplying us the numerical data of ref. \citen{ak}. The numerical calculation is performed using the program package TITPACK version 2 coded by H. Nishimori on HITAC S820 and SR2201 at the Information Processing Center of Saitama University and the FACOM VPP500 at the Supercomputer Center of Institute for Solid State Physics, University of Tokyo. This work is supported by a Grant-in-Aid for Scientific Research from the Ministry of Education, Science, Sports and Culture of Japan and a research grant from the Natural Science and Engineering Research Council of Canada (NSERC).

\end{document}